          [2001/04/25 1.1 (PWD)]
\documentclass[a4paper,twocolumn]{esapub2005} 
\usepackage{times}
\usepackage{graphicx,natbib,epsfig}

\title{MHD versus kinetic effects in the solar coronal heating: a two stage mechanism}
\author{David Tsiklauri}
\affil{Institute for Materials Research,
University of Salford, Greater Manchester, M5 4WT, United Kingdom}

\begin{document}


\maketitle

\begin{abstract}
Using Particle-In-Cell simulations i.e. in the kinetic plasma description the 
discovery of a new mechanism of parallel electric field generation was recently 
reported. Here we show that the electric field generation parallel to the uniform 
unperturbed magnetic field  can be obtained in a much simpler framework using the 
ideal magnetohydrodynamics (MHD) description. In ideal MHD the electric field 
parallel to the uniform unperturbed magnetic field  appears due to fast magnetosonic 
waves which are generated by the interaction of weakly non-linear Alfv\'en waves 
with the transverse density inhomogeneity. In the context of the coronal heating 
problem  a new {\it two stage mechanism} of plasma heating is presented by putting 
emphasis, first, on the generation of parallel electric fields within an {\it ideal MHD} 
description directly,  rather than focusing on the enhanced dissipation mechanisms 
of the Alfv\'en waves and, second, dissipation of these parallel electric fields 
via {\it kinetic} effects.  It is shown that for a single 
 Alfv\'en wave harmonic  with frequency $\nu = 7$ Hz, and longitudinal wavelength  
 $\lambda_A = 0.63$ Mm for a putative  Alfv\'en speed of 4328 km s$^{-1}$,
the generated parallel electric field could account for 10\% of the necessary 
coronal heating requirement. We conjecture that  wide spectrum (10$^{-4}-10^3$ Hz) 
Alfv\'en waves, based on the observationally constrained spectrum, could provide 
the necessary coronal heating requirement. By comparing MHD versus kinetic results 
we also show that there is a clear indication of the {\it anomalous resistivity} which 
is 100s of times greater than the classical Braginskii value.
\end{abstract}

\section{Introduction and Motivation}

The coronal heating problem, the puzzle of what maintains the solar corona 200 times
hotter than the photosphere, is one of the main outstanding questions in solar physics.
A significant amount of work has been done in the context of heating of open
magnetic structures in the solar 
corona (e.g. phase mixing, one of the possible mechanisms of the heating).
Historically all phase mixing studies have been performed in the Magnetohydrodynamic (MHD) approximation,
however, since the transverse scales in the Alfv\'en wave collapse progressively to zero,
the MHD approximation is inevitably violated. 
Thus, \citet{tss05a,tss05b} studied the phase mixing effect in the kinetic regime, using Particle-In-Cell simulations, i.e.
beyond a MHD approximation, where
a new mechanism for the acceleration of electrons
due to the generation of a parallel electric field in the solar coronal context was discovered. 
This mechanism has important implications
for various space and laboratory plasmas, e.g. the 
coronal heating problem and acceleration of the solar wind.
It turns out that in the magnetospheric context, a similar parallel electric field generation 
mechanism in the transversely inhomogeneous plasmas has been previously reported 
\citep{glm04,glq99}. See also \citet{mgl06} and references therein.

After the comment paper by \citet{mgl06} we came to the realisation that 
the electron acceleration seen in both series of 
works \citep{tss05a,tss05b,glm04,glq99} is a non-resonant wave-particle 
interaction effect. In works by \citet{tss05a,tss05b}
the electron thermal speed was $v_{th,e}=0.1c$ while the Alfv\'en speed 
in the strongest density gradient regions
was $v_A=0.16c$; this unfortunate coincidence led us to the conclusion that the electron acceleration by parallel
electric fields was affected by the Landau resonance with the 
phase-mixed Alfv\'en wave. In works by \citet{glm04,glq99}
the electron thermal speed was $v_{th,e}=0.1c$ while the Alfv\'en speed was $v_A=0.4c$ because they considered a more
strongly magnetised plasma applicable to Earth magnetospheric conditions. 
However, the interaction of the Alfv\'en wave with a transverse density plasma inhomogeneity
when the Landau resonance condition $\omega=k_\parallel c_A(x)$ is met can be quite important for the electron acceleration
\citep{chaston2000,hs76}. \citet{chaston2000} assert that the electron 
acceleration observed in density cavities in aurorae
can be explained by the Landau resonance of the cold ionospheric 
electrons with the Alfv\'en wave.
\citet{hs76} also established that at the resonance  
Alfv\'en wave fully converts into the kinetic  Alfv\'en wave with
the perpendicular wavelength comparable to the ion gyro-radius. 
We can then conjecture that {\it because of  kinetic Alfv\'en wave front
stretching} (due to phase mixing, i.e. due to the differences in local Alfv\'en speed), 
this perpendicular component {\it gradually realigns} with the ambient 
magnetic field and hence
creates the time varying parallel electric field component. 
This points to the importance of the Landau resonance 
for electron acceleration when the resonance condition is met. 
But as witnessed from works of  \citet{glm04,glq99},
even when the resonance condition is not met electron acceleration is still possible.

\begin{figure*}
\centering
\epsfig{file=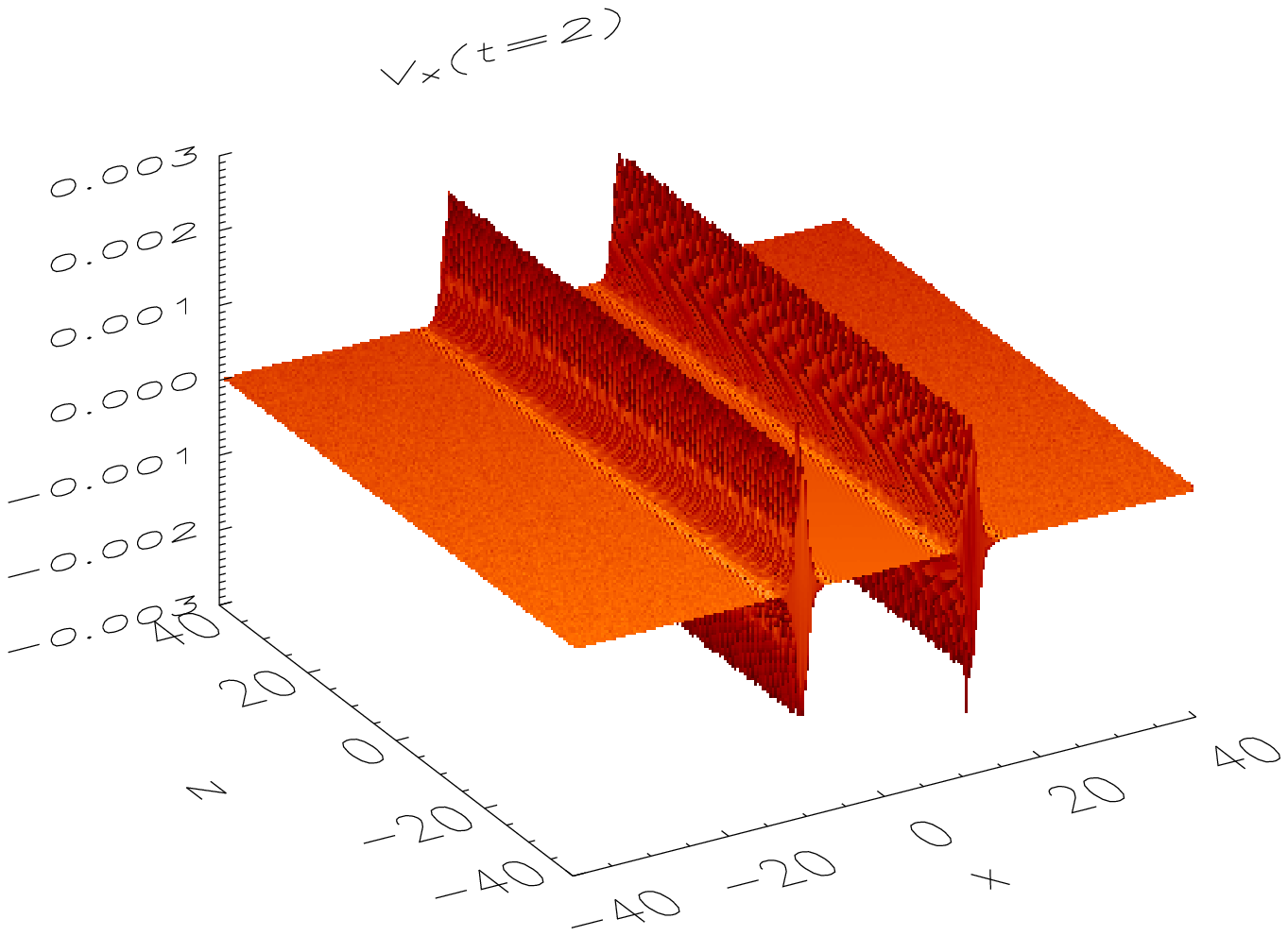,width=6cm}
 \epsfig{file=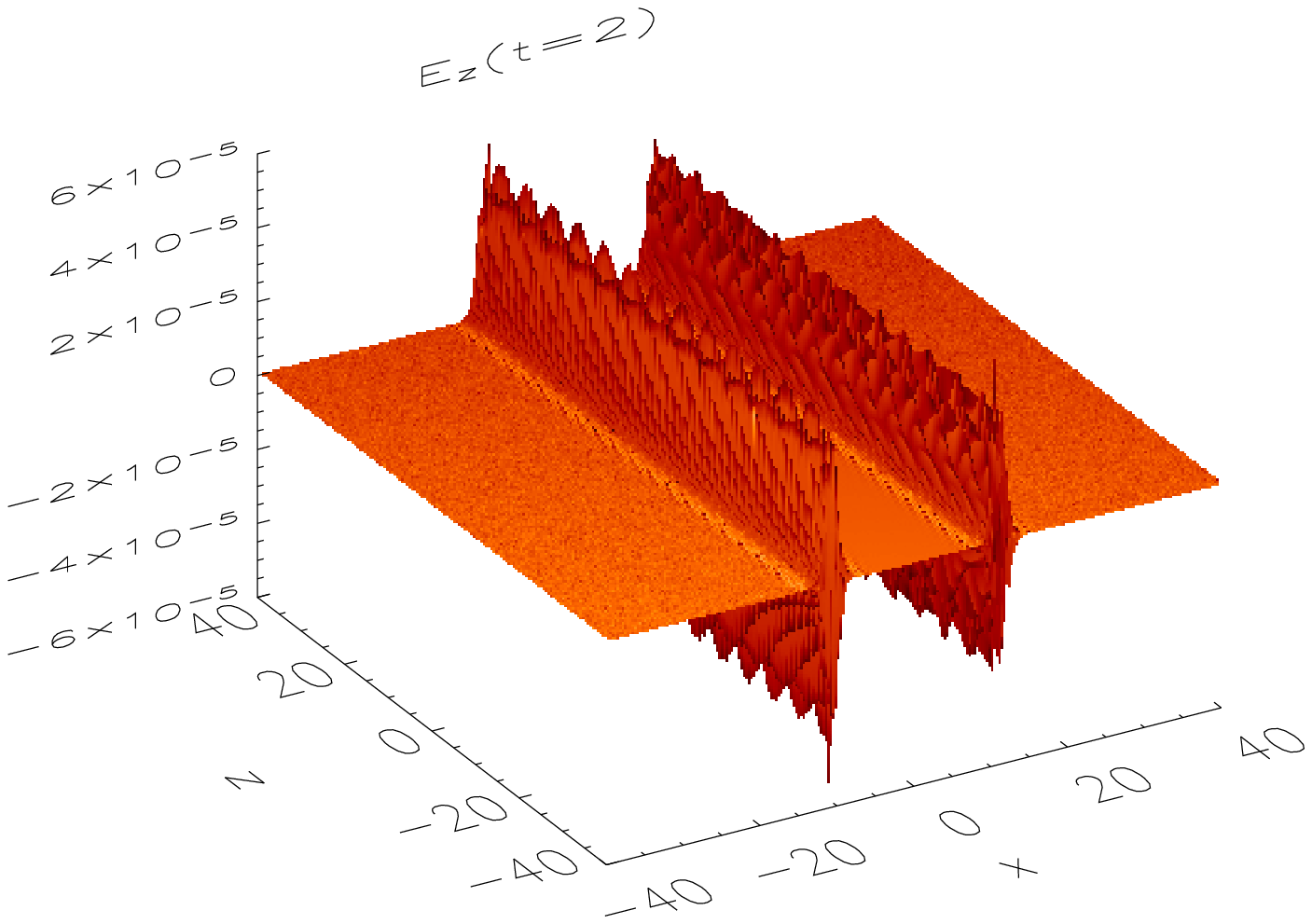,width=6cm}
 \epsfig{file=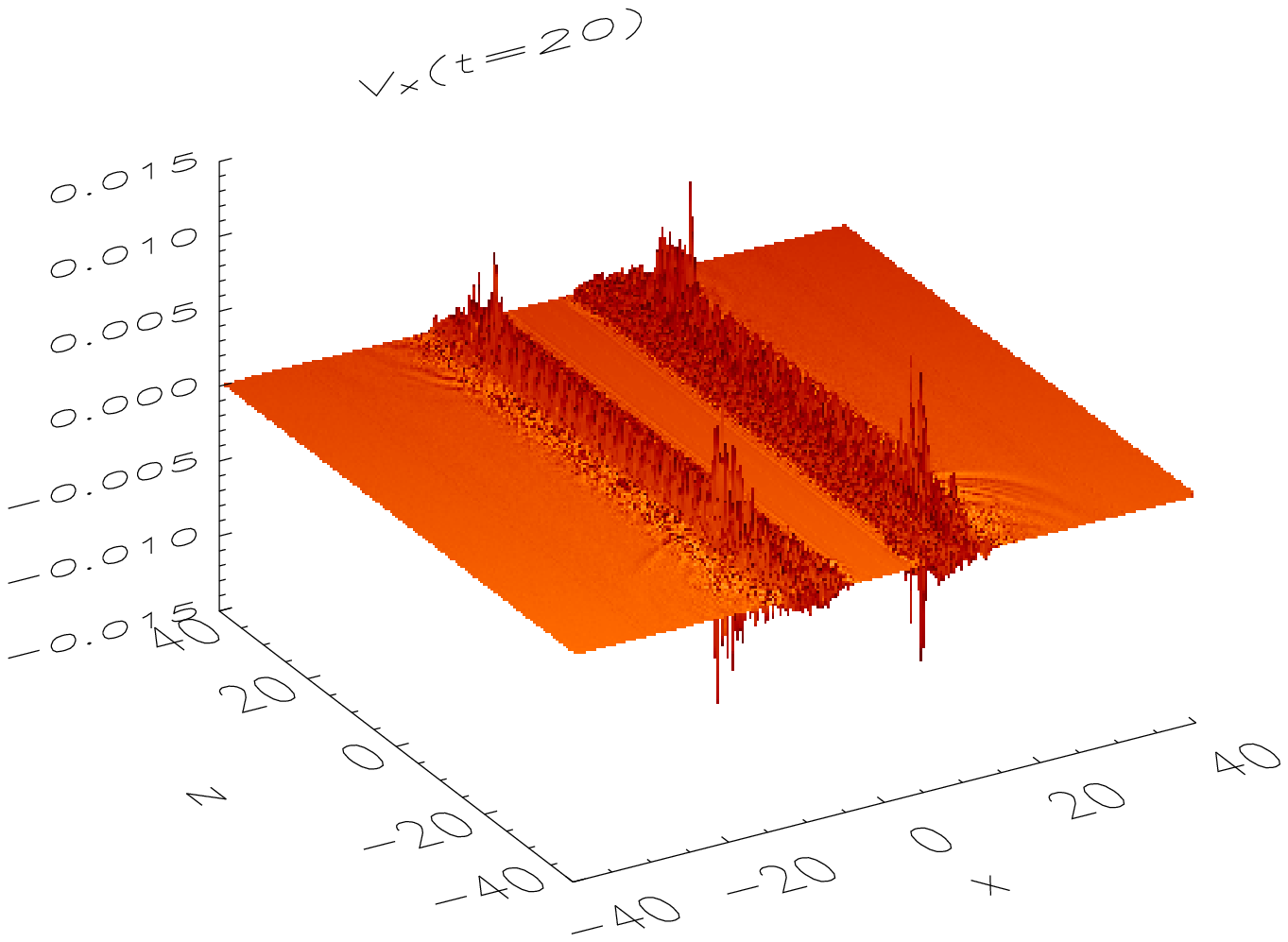,width=6cm}
 \epsfig{file=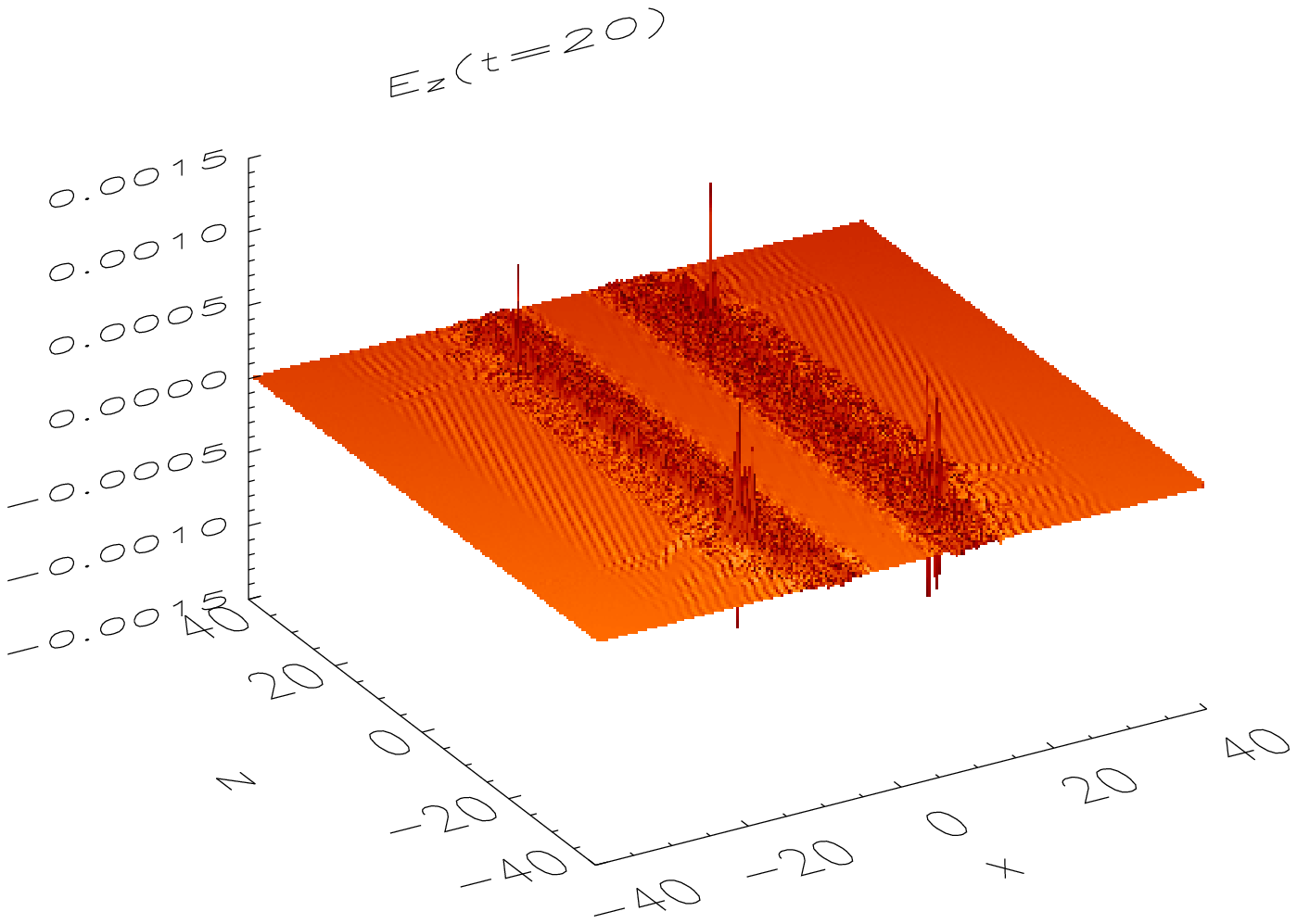,width=6cm}
 \caption{Snapshots of $V_x$ and $E_z$ at $t=2$ and 20 for the case of $k=10$, $\nu = 7$ Hz, $\lambda_A = 0.63$ Mm.}
\end{figure*}

There were three main stages that lead to the formulation of the present model.

(i) The realisation of the parallel electric field generation (and particle acceleration) being
 a {\it non-resonant} wave-particle interaction effect lead us to the question: 
 could such  parallel electric fields be generated in a MHD approximation? 

(ii) Next we realised that if one considers
{\it non-linear} generation of the fast magnetosonic waves in the transversely inhomogeneous plasma, 
then $\vec E = - (\vec V \times \vec B)/c$ contains a non-zero component
parallel to the ambient magnetic field $E_z = - (V_x B_y - V_y B_x)/c$.  

(iii) From previous studies \citep{bank00,tan01} we knew that the fast magnetosonic 
waves ($V_x$ and $B_x$) did not grow to
a substantial fraction of the Alfv\'en wave amplitude. However after reproducing the old 
parameter regime (k=1, i.e a frequency of 0.7 Hz),
the case of k=10, i.e a frequency of 7 Hz was considered, which showed that  fast magnetosonic waves
and in turn a parallel electric field were more efficiently generated.

\section{model, rationale,  and main results}

Unlike previous studies \citep{tss05a,tss05b,glm04,glq99}, here we use an ideal MHD description of the problem.
We solve numerically ideal, 2.5D, MHD equations in Cartesian coordinates, 
with a plasma beta of 0.0001 starting from the following equilibrium
configuration: A uniform magnetic field $B_0$ in the $z-$direction penetrates plasma with the density
inhomogeneity across the $x-$direction, which varies according to
\begin{equation}
\rho(x)=\rho_0\left[1+2\left(\tanh(x+10)+\tanh(-x+10)\right)\right]. 
\end{equation}
This  means that the plasma density increases from 
some reference background value of $\rho_0$,
which in our case was fixed at 
$\rho_0=2\times10^9 \mu m_p$ g cm$^{-3}$ (with a molecular weight of 
$\mu=1.27$ corresponding to the solar coronal conditions $^1$H:$^4$He=10:1 
and $m_p$ being the proton mass), to $5 \rho_0$.
Such a density profile across the magnetic field has steep gradients with a half-width of 3 Mm around
$x \simeq \pm 10$ Mm and is essentially flat elsewhere. Such a structure mimics e.g.
the footpoint of a large curvature radius solar coronal loop or a polar
region plume  with the ratio of the density inhomogeneity scale and the loop/plume radius of 0.3, which is
the median value of the observed range 0.15 - 0.5.

The initial conditions for the numerical simulation are
$B_y=A \cos(kz)$ and $V_y=-c_A(x)B_y$ at $t=0$, which means that a purely 
Alfv\'enic, linearly polarised, plane wave is launched travelling in the
direction of positive $z$s. The rest of the physical quantities, 
 $V_x$ and $B_x$ (which 
would be components of fast magnetosonic waves if the medium were totally homogeneous)
and $V_z$ and $B_z$ (the analogs of slow magnetosonic waves) are initially set
to zero. 
The plasma temperature is varied as the inverse of Eq.(1) so that the total pressure always
remains constant.  Boundary conditions used in our simulations are periodic along the $z$- and the zero gradient along the $x$-coordinates.
We fixed the amplitude of  Alfv\'en wave $A$ at 0.05 throughout. This choice makes
the Alfv\'en wave weakly non-linear.

As a self-consistency test, we considered situation when the wavenumber of the initial
Alfv\'en wave is $k=1$.  In dimensional units this corresponds to an 
Alfv\'en wave with frequency ($\nu = 0.7$ Hz), i.e. longitudinal wave-numbers $\lambda_A = 6.3$ Mm.
We corroborated the previous results of \citet{bank00} and \citet{tan01}.

We now present results for the case of large wave-numbers, $k=10$,   which in dimensional units 
correspond to an Alfv\'en wave with $\nu = 7$ Hz, and $\lambda_A = 0.63$ Mm.
This is a regime not investigated before.
In Fig.~(1) we show shaded surface plots of both the fast magnetosonic waves ($V_x$) and parallel electric field $E_z$. 
We gather from this graph that similarly to the results of 
\citet{tss05b} and \citet{glm04}, the generated parallel electric field
is quite spiky, but more importantly  
large wave-numbers i.e. short wavelength now are able to
significantly increase the amplitudes of 
both the fast magnetosonic waves ($V_x$) and parallel electric field $E_z$.
This amplitude 
growth is beyond a simple $A^2$ scaling.
 The amplitude growth is presented quantitatively in Fig.~(2) (by doubling spatial resolution
 we perform satisfactory convergence test).
The amplitude of  $V_x$ now attains values of 0.01 unlike for moderate $k$s.
Thus, large wavenumbers (i.e. stronger spatial gradients) seem to 
yield larger values for the level of saturation of the $V_x$ amplitude.
In the considered case, $E_z$ now attains values of 0.001.

The coronal energy losses that need to be compensated
by some additional energy input, to keep the solar corona at the observed temperatures, are  (in units of erg cm$^{-2}$ s$^{-1}$):
$3\times10^5$ for the quiet Sun, $8\times10^5$ for a coronal hole and $10^7$ for an active region.
One can estimate the heating flux per unit area (i.e. in erg cm$^{-2}$ s$^{-1}$):
\begin{equation}
F_H=E_H \lambda_T=5 \times 10^3 \left(\frac{n_e}{10^8 {\rm cm}}\right)^2\left(\frac{T}{1 {\rm MK}}\right),
\end{equation}
where $E_H \approx 10^{-6}$ erg cm$^{-3}$ s$^{-1}$. This yields an estimate of $F_H \approx 2\times10^6$ erg cm$^{-2}$ s$^{-1}$
in an active region with a typical loop base electron number density of $n_e=2\times10^9$ cm$^{-3}$ and $T=1$ MK.

The energy density associated
with the parallel electric field $E_z$ is
\begin{equation}
E_E=\frac{\varepsilon E_z^2} {8 \pi} ,
 \;\;\;
\left[{\rm erg \, cm^{-3} }\right]
\end{equation}
where $\varepsilon$ is the dielectric permitivity of plasma. The latter can be deduced
from 
\begin{equation}
\varepsilon=\frac{4 \pi \rho c^2}{B^2}.
\end{equation}
For the coronal conditions ($\rho=2\times10^9 \mu m_p$ g cm$^{-3}$, 
$\mu=1.27$, $B=100$ Gauss) $\varepsilon \approx  4.8048 \times 10^3 $.
In Fig.~(2) we saw that electric field amplitude attains a value of $\approx 0.001$. In order to convert
this to dimensional units we use $c_A^0=4328$ km s$^{-1}$ and $B=100$ G and $E_z = - (V_x B_y - V_y B_x)/c$ to obtain
$E_z\approx(c_A^0 B/c)\times 0.001=0.0014$ statvolt cm$^{-1}$ (in Gaussian units).
Therefore the energy density associated
with the parallel electric field $E_z$ (From Eq.(3)) is 
\begin{equation}
E_E=\varepsilon \times 0.0014^2/(8 \pi)=3.7471 \times 10^{-4}
\end{equation}
$\left[{\rm erg \, cm^{-3} }\right].$

In order to get  the heating flux per unit area for a {\it single harmonic} with frequency
7 Hz, we multiply the latter expression by the 
Alfv\'en speed of 4328 km s$^{-1}$  to obtain
\begin{equation}
F_E=E_E c_A^0=1.62\times10^5\;\;\;
\left[{\rm erg \, cm^{-2} s^{-1}}\right],
\end{equation}
which is $\approx 10$ \% of the
coronal heating requirement estimate for the same parameters made above using Eq.(2).
Note that the latter estimate is for  a {\it single harmonic} with frequency
7 Hz.

A crucial next step that is needed to understand how the generated electric fields parallel to the uniform unperturbed magnetic field
 dissipate {\it must invoke kinetic effects}. In our two stage model, in the first stage bulk MHD motions (waves) 
generate the parallel electric fields, which cannot accelerate particles if we describe plasmas in the ideal MHD
limit. \citet{glm04} and \citet{tss05b} showed that when the identical system is
modelled in the kinetic regime particles {\it are} accelerated with such parallel fields and Alfv\'en wave energy
is converted into heat on a time scale of a few Alfv\'en periods.

\begin{figure}[]
\resizebox{\hsize}{!}{\includegraphics{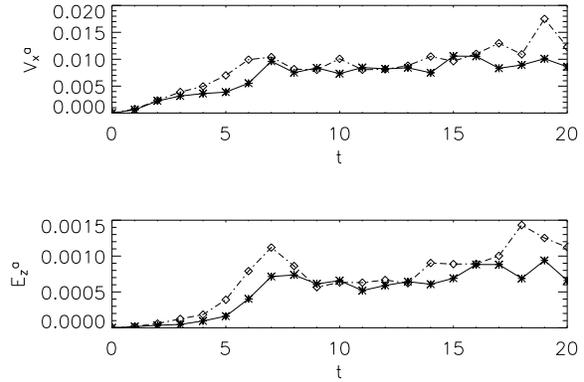}} 
\caption{Time evolution of the amplitudes of $V_x \equiv V_x^a$ and $E_z \equiv E_z^a$.
Solid lines with stars represent solutions using the Lare2d code with $4000 \times 4000$ resolution, while
dash-dotted lines with open symbols are the same but with $2000 \times 2000$ resolution. 
Here $A=0.05$, $k=10$, $\nu = 7$ Hz, $\lambda_A = 0.63$ Mm.}
\end{figure}

Alfv\'en waves as observed in situ in the solar wind always appear to be
propagating away from the Sun and it is therefore natural to assume a solar origin for
these fluctuations. However, the precise origin in the solar atmosphere of the hypothetical
source spectrum for Alfv\'en waves (turbulence) is unknown, given the impossibility of remote
magnetic field observations above the chromosphere-corona transition region.
Studies of ion cyclotron resonance heating of the solar corona and high speed winds exist which
provide important spectroscopic constraints on the Alfv\'en wave spectrum.
Although the spectrum can and is observed at distances of 0.3 AU, it can be projected back to the base of
corona using empirical constraints. Therefore, we conjecture that  wide spectrum (10$^{-4}-10^3$ Hz) Alfv\'en waves, based on the 
observationally constrained spectrum, could
provide the necessary coronal heating requirement.
The exact amount of energy that could be deposited by such waves through our mechanism of parallel electric field generation can
only be calculated once a more complete parametric study is done. Thus, the "theoretical spectrum" of the energy stored in 
parallel electric fields versus frequency needs to be obtained. 
At present we only have two points, 0.7 Hz and 7 Hz, in our "theoretical spectrum".
Preliminary results will be presented elsewhere \citep{t06a,t06b}.

\section{MHD versus kinetic effects and anomalous resistivity}
There are two main candidates for the solution of the coronal heating problem:
so-called DC and AC models. DC or magnetic reconnection based 
models need to invoke anomalous resistivity (somewhat ad hoc concept, but there is 
a lot of indirect evidence for it). At present the  details of reconnection in 3D are not understood in full.
AC models are often not quantitative, or not enough heating can be provided, 
unless some enhanced dissipation mechanisms (e.g. phase-mixing, resonant absorption) are invoked.
This makes a perfect playground for studying interplay between MHD and kinetic theories of the solar corona.
The reason is two-fold: 
1) The issue of anomalous resistivity on which DC models rely can 
only be settled by studying kinetic effects (micro-physics);
2) AC models which use MHD eventually break down, as often 
system naturally evolves towards progressively small scales (e.g. 
in phase-mixing); Therefore, kinetic effects become important.

\cite{tss05a,tss05b}, amongst other findings, established that the Alfv\'en wave 
amplitude decay law in the inhomogeneous regions, 
in the kinetic regime is 
 $\propto \exp \left[ -
\left({x \Delta}/{1250 \Delta}\right)^3\right]$ (where $x$ is the coordinate along uniform magnetic field -- mind the change of the geometry!);
which is the same as in the MHD approximation 
discovered by Heyvaerts and Priest (1983) (HP83 thereafter):
$\propto \exp \left[ -
\left(\frac{\eta \omega^2 V_A^{\prime 2}}{6 V_A^{5}}\right)x^3\right]$.
Question that begs to be asked: what if we calculate resistivity 
from the latter MHD formula using our kinetic (PIC) empirical 
dissipation length of $1250\Delta$?  (F. Malara, private communication).
By equalising expressions under the both exponents and noting that $x$ from HP83 is equal to $x\Delta$ from
our kinetic formula, we obtain $\eta=6 V_A^5/[\omega^2 V_A^{\prime 2}(1250 \Delta)^3] $.
In order to estimate $V_A^{\prime}=dV_A/dy$ we put $V_A^{\prime}\approx V_A(y_*)/\delta y$ where $V_A(y_*)$ is
the Alfv\'en speed at the strongest gradient point and $1/\delta y$ is the strength of the gradient.
The latter can be approximated as $(V_A(\infty)-V_A(0)) \delta t / (50\delta) \cdot 1/(50 \delta)$, with $50 \delta$
being the scale of the transverse density gradient. Putting $\omega=0.3\omega_{ci}$ and $\delta t=54.69 / \omega_{ci}$
we obtain $\eta=7\times10^{-5} V_A^3(y_*) \Delta/(V_A(\infty)-V_A(0))^2$.
From \cite{tss05a,tss05b} $V_A(y_*)=0.16c$ and $(V_A(\infty)-V_A(0))=0.25c-0.125c=0.125c$, 
hence $\eta=1.8\times10^{-5}c\Delta$. Also, $\Delta=v_{th,e}/\omega_{ce}$. 
The thermal speed of electrons (at infinity) $v_{th,e}=\sqrt{kT/m_e}$, which for 1MK coronal temperature 
is $3.9\times 10^8$ cm s$^{-1}$.
The electron cyclotron frequency $\omega_{ce}=1.76\times10^7B[G]$, which for 10 G 
magnetic field is $1.8\times 10^8$ s$^{-1}$. Thus, $\Delta=2$ cm and we finally obtain $\eta=1.1\times10^6$ cm$^2$ s$^{-1}$,
or in SI units $\eta=1.1\times10^2$ m$^2$ s$^{-1}$. This is about 100 times greater than the classical Braginskii value
for the resistivity. Therefore, the obtained value is a clear indication of the {\it anomalous resistivity}. Interestingly others quote similar values for
the $\eta$ \citep{pwh03,whf02}.

\section*{Acknowledgments}

Author kindly acknowledges support from the Nuffield Foundation (UK) through an award to newly 
appointed lecturers in Science, Engineering and Mathematics (NUF-NAL 04); from the 
University of Salford Research Investment Fund 2005 grant; and 
use of the E. Copson Math cluster funded by PPARC 
and the University of St. Andrews.


\end{document}